\documentclass[sigconf]{acmart}

\usepackage{booktabs} 
\usepackage{multirow}
\usepackage{makecell}

\setcopyright{rightsretained}

\acmDOI{}

\acmISBN{}

\acmConference[DLfM'17]{4th International Digital Libraries for Musicology workshop}{October 2017}{Shanghai, China} 
\acmYear{2017}
\copyrightyear{2017}

\acmPrice{15.00}

\begin{document}
\title{Creating an A Cappella Singing Audio Dataset \\for Automatic Jingju Singing Evaluation Research }

\author{Rong Gong}
\affiliation{%
  \institution{Music Technology Group\\ Universitat Pompeu Fabra}
  \city{Barcelona} 
  \state{Spain} 
}
\email{rong.gong@upf.edu}

\author{Rafael Caro Repetto}
\affiliation{%
  \institution{Music Technology Group\\ Universitat Pompeu Fabra}
  \city{Barcelona} 
  \state{Spain} 
}
\email{rafael.caro@upf.edu}

\author{Xavier Serra}
\affiliation{%
  \institution{Music Technology Group\\ Universitat Pompeu Fabra}
  \city{Barcelona} 
  \state{Spain} 
}
\email{xavier.serra@upf.edu}

\renewcommand{\shortauthors}{R. Gong et al.}

\begin{abstract}
The data-driven computational research on automatic jingju (also known as Beijing or Peking opera) singing evaluation lacks a suitable and comprehensive a cappella singing audio dataset. In this work, we present an a cappella singing audio dataset which consists of 120 arias, accounting for 1265 melodic lines. This dataset is also an extension our existing CompMusic jingju corpus. Both professional and amateur singers were invited to the dataset recording sessions, and the most common jingju musical elements have been covered. This dataset is also accompanied by metadata per aria and melodic line annotated for automatic singing evaluation research purpose. All the gathered data is openly available online\footnote{https://doi.org/10.5281/zenodo.842229}.
\end{abstract}

%
%



\maketitle

\section{Introduction}\label{sec:intro}

\subsection{Short presentation of jingju music}
The music of jingju has been receiving increasing attention from MIR researchers in the last years. A brief jingju MIR research bibliography can be referred to \cite{Repetto2017Score}. Music in jingju has been deeply conventionalized according to the following three elements that build its musical system:
\begin{itemize}
    \item \textit{shengqiang}: melodic framework associated with a particular emotional atmosphere. There are two main \textit{shengqiang}, namely \textit{xipi} and \textit{erhuang}, which define the musical identity of jingju.
    \item \textit{banshi}: rhythmic transformations of the \textit{shengqiang}´s melodic framework, which can be classified into two categories: metered and non-metered. 
    \item role-type: acting profile which the performer belongs to. There are four broad categories of role-types: \textit{sheng}, \textit{dan}, \textit{jing} and \textit{chou}, where \textit{chou} is focused on other disciplines than singing like reciting, acting or acrobatics. For the role-types focused on singing, their style is a variation of either the male or female style, represented respectively by \textit{laosheng} and \textit{dan}.
\end{itemize}
The structure of the lyrics determines the musical structure of the arias. The basic lyric unit for jingju arias is the couplet, and each \textit{shengqiang} defines a melodic line for the opening line of the couplet, and another one for the closing line. One single aria is usually set to only one \textit{shengqiang}, but it can contain different \textit{banshi}.

\subsection{Automatic jingju singing evaluation}
\begin{figure}[!ht]
\includegraphics[scale=0.6]{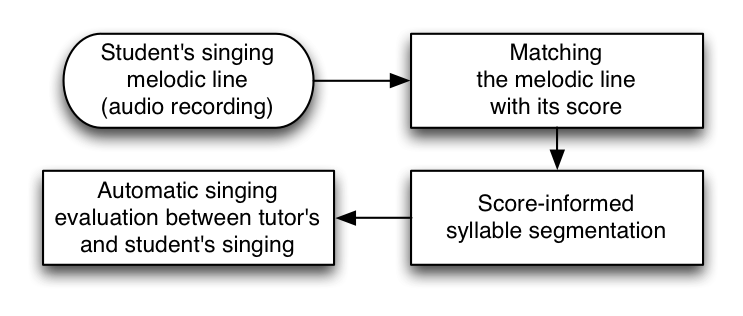}
\caption{Flow chart of the research project.}
\label{fig:system_framework}
\end{figure}


The most prominent aspect of jingju music is singing. The ultimate goal of our research project is to automatically evaluate the jingju a cappella singing of a student in the scenario of jingju singing education, see figure \ref{fig:system_framework}. Jingju is extremely demanding in the clear pronunciation and accurate intonation for each syllabic or phonetic singing unit. During the initial learning stages, students are required to imitate completely tutor's singing. Therefore, the automatic jingju singing evaluation system we envision is based on this training principle and measures the intonation and pronunciation similarities between the student's and the tutor's melodic lines. Before measuring the similarities, the a cappella singing melodic lines should be matched with their scores \cite{Rong2017Matching}; then the score-informed method will be used to segment these lines into syllabic or phonetic units in order to capture the temporal details \cite{Jordi2017Segmenting}. Considering that our research mainly uses data-driven methods, it is thus necessary to build a relatively large a cappella singing audio dataset in order to better train and validate the computational models for automatic jingju singing evaluation. 

\subsection{Existing jingju music audio datasets}\label{sec:existing datasets}
The CompMusic corpus \cite{Repetto2014Creating} is formed by a collection of commercial recordings, as well as their metadata. Another jingju music corpus gathered in \cite{Tian2016TMS} also consists of commercial recordings, annotated for structural segmentation analysis. These recordings are all mixed with instrumental accompaniment, which means clean singing voice should be separated during the preprocessing step if we want to take advantage of these recordings for the automatic singing evaluation research. However, singing voice separation itself is a very challenging research task. One of the state-of-the-art source separation algorithms \cite{chandna2017monoaural} was tried by the authors, which was not able to produce a clean singing voice while preserving its characteristics.  

The only jingju a cappella singing dataset we can find \cite{black2014automatic} is small and not complete enough since it only contains around 1 hour recordings of 31 unique arias, and its annotations were made for the task of mood recognition instead of automatic singing evaluation. This dataset has been used intensively in our previous studies for the subtasks of pitch contour segmentation \cite{gong2016pitch}, acoustic modeling of phonemes \cite{Rong2017Matching} and syllable segmentation \cite{gong2017score,Jordi2017Segmenting}. The results of these studies revealed that such small dataset is the main bottleneck which prevents building robust computational models with better generalization ability.  

The main goal of the dataset presented in this paper is to offer a comprehensive and complete resource for the study of automatic jingju singing evaluation. The recordings of this dataset show the most common \textit{shengqiang}, \textit{banshi} and role-types in the jingju singing educational scenario. Both professional and amateur recordings were collected so that similarity or evaluation models can be built between them. The remaining of the paper is structured as follows. In the next section the jingju a cappella singing audio dataset is described in detail. 
In the last section, we present some concluding remarks and point out future work.

\section{Recording the dataset}

\subsection{Artists}
 We invited 5 professional singers from NACTA (National Academy of Chinese Theatre Arts, all of them have rich experience in stage performance and teaching) and another 4 amateur singers from jingju associations in non-art schools to the recording sessions. All singers were asked to sing their familiar arias with the constraint that these arias should be representative ones for each school.

Jingju singing is accompanied by an instrumental ensemble including minimum 8 melodic and percussive instruments in heterophony. It is customary that when it is short of instrument players or in the scenario of singing practice, only the primary instrument of the ensemble - \textit{jinghu}, is used as the accompaniment because the other melodic instruments follow the melody played by the \textit{jinghu} \cite{wichmann1991listening}. 7 singers (3 professional and 4 amateur) were singing along with the accompaniment of commercial audio recordings; other 2 professional singers were accompanied by 2 professional \textit{jinghu} players. We show the detail information of the singers and \textit{jinghu} players in table \ref{tab:artists}.

\subsection{Recording setup}
Most of the recording sessions have been conducted in professional recording rooms by using professional equipment, see table \ref{tab:equipment_room} for the detail information, where we use two recording equipment sets and two recording rooms:
\begin{itemize}
    \item Set 1: M-Audio Luna condenser microphone + RME Fireface UCX audio interface + Apple GarageBand for Mac DAW;
    \item Set 2: Mojave MA-200 condenser microphone + ART voice channel microphone preamp + RME Fireface 800 audio interface + Adobe Audition 2.0 DAW;
    \item Room 1: The conference room in NACTA's business incubator with reflective walls, carpet-covered floor, conference furniture and medium room reverberation;
    \item Room 2: The sound recording studio in Institute of Automation, Chinese Academy of Science, with acoustic absorption and isolation.
\end{itemize}

\begin{table}[!ht]
\caption{Equipment and rooms for recording each artist.}
\label{tab:equipment_room}
\begin{tabular}{lll}
\toprule
\makecell[l]{Recording \\equipment} & \makecell[l]{Recording \\rooms} & Artists' names                              \\
\midrule
set 1                & room 1  &   LIAO Jiani                                                  \\
\multicolumn{2}{c}{ \makecell[l]{Partly with set 1 in room 1 \\ and
partly with set 2 in room 2} } & SHAO Yakun   \\
set 2                & room 2    & \makecell[l]{SUN Yuzhu, SONG Ruoxuan, \\TIAN Hao, LONG Tianming,\\ LIU Hailin, SONG Weihao,\\ XU Jingwei, ZHANG Lantian,\\ FU Yanchen}                                                \\
\bottomrule
\end{tabular}
\end{table}

When commercial audio recordings were used as the accompaniment, singers were recorded while listening to the accompaniment sent through their monitoring headphone. Otherwise, when \textit{jinghu} players were used as the accompaniment, to simultaneously record both singing and \textit{jinghu} without crosstalk, we placed them separately in two different recording rooms and used two recording channels. However, they were still able to have visual communication through a window and monitor each other through headphones.

\section{Description of the dataset}
In total, 21 recording sessions were conducted, which resulted in a dataset containing around 9 hours audio recordings of 120 arias - 74 of them are sung by professional singers and 46 are sung by amateur singers. Table \ref{tab:works} shows the distribution of aria recordings per role-type and \textit{shengqiang}; \textit{banshi} is not included here because some arias contain more than one. However, since the main melodic unit is the line, the information given in Table \ref{tab:lines} is a better representation of the dataset's potential for the study of the automatic jingju singing evaluation.

\begin{table*}[t]
\caption{Detail information of the recording artists, according to role-type, singing school, affiliation, level and use of accompaniment. NACTA: National Academy of Chinese Theatre Arts, USTB: University of Science and Technology Beijing, Renmin: Renmin University of China}
\label{tab:artists}
\begin{tabular}{llllllll}
\toprule
                                  &                                 & Artist's name & Role-type & Singing school                   & Affiliation              & Level     & \makecell[l]{Use of \\ accompaniment}   \\
\midrule
\multirow{9}{*}{Singers}        & \multirow{5}{*}{ Professional } & SUN Yuzhu (female)     & \textit{dan}       & CHENG Yanqiu             & \multirow{5}{*}{NACTA} & \makecell[l]{4th year \\ undergraduate}  & audio recordings       \\
                                  &                                 & LIAO Jiani (male)    & \textit{laosheng}  & \makecell[l]{YU Shuyan \& \\ YANG Baoseng} &                          & \makecell[l]{4th year \\ undergraduate}  & audio recordings       \\
                                  &                                 & SHAO Yakun (female)    & \textit{jing}      & QIU Shengrong            &                          & \makecell[l]{4th year \\ undergraduate}   & audio recordings       \\
                                  &                                 & SONG Ruoxuan (female)  & \textit{dan}       & MEI Lanfang              &                          & graduated & FU Yanchen             \\
                                  &                                 & TIAN Hao (male)      & \textit{laosheng}  & YU Shuyan                &                          & graduated & ZHANG Lantian          \\ \cline{2-8}
                                  & \multirow{4}{*}{ Amateur }      & LIU Hailin (male)    & \textit{dan}       & CHENG Yanqiu             & USTB                     & --        & audio recordings                  \\
                                  &                                 & SONG Weihao (male)   & \textit{dan}       & XUN Huisheng             & USTB                     & --        & audio recordings                 \\
                                  &                                 & LONG Tianming (male) & \textit{laosheng}  & \makecell[l]{YU Shuyan \& \\ YANG Baoseng} & USTB                     & --        & audio recordings                 \\
                                  &                                 & XU Jingwei (male)    & \textit{laosheng}  & --                       & Renmin                   & --        & audio recordings                 \\ \hline
\multirow{2}{*}{\makecell[l]{Jinghu\\players}} & \multirow{2}{*}{ Professional } & ZHANG Lantian & --        & --                       & \multirow{2}{*}{NACTA} & \makecell[l]{3rd year \\ undergraduate}  & --                     \\
                                  &                                 & FU Yanchen    & --        & --                       &                          & graduated & --                     \\
\bottomrule
\end{tabular}
\end{table*}

\begin{table}[!ht]
\caption{Content of the jingju a cappella audio dataset, according to role-type and \textit{shengqiang}, Format of each cell: professional\textbar amateur aria number}
\label{tab:works}
\begin{tabular}{lcccc}
\toprule
              & \textit{laosheng} & \textit{dan}   & \textit{jing} & Total   \\
\midrule
\textit{xipi}          & 16\textbar17    & 19\textbar4  & 10\textbar0 & 45\textbar21   \\
\textit{erhuang}       & 13\textbar10    & 4\textbar8   & 4\textbar0  & 21\textbar18   \\
\textit{sipingdiao}    & --       & 4\textbar2   & --   & 4\textbar2     \\
\textit{nanbangzi}     & --       & 2\textbar1   & --   & 2\textbar1     \\
\textit{fanerhuang}    & 0\textbar2      & 1\textbar1   & --   & 1\textbar3     \\
\textit{fansipingdiao} & --       & 1\textbar1   & --   & 1\textbar1     \\
Total         & 29\textbar29    & 31\textbar17 & 14\textbar0 & 74\textbar46   \\
\bottomrule
\end{tabular}
\end{table}

\begin{table*}[t]
\caption{Content of the jingju a cappella audio dataset per melodic line for role-types of \textit{dan}, \textit{laosheng} and \textit{jing}, according to \textit{shengqiang} and \textit{banshi}. On the upper heading, \textit{da} stands for \textit{dan}, \textit{ls} for \textit{laosheng}, \textit{eh} for \textit{erhuang} and \textit{xp} for \textit{xipi}. Format of each cell: professional\textbar amateur melodic line number (related lines in the score collection \cite{Repetto2017Score}).}
\label{tab:lines}
\begin{tabular}{llllllll}
\toprule
            & \textit{daeh}        & \textit{daxp}           & \textit{lseh}            & \textit{lsxp}            & \textit{jieh} & \textit{jixp}  & Total               \\
\midrule
\textit{yuanban}     & 0\textbar27        & 58\textbar22 (20\textbar22)  & 54\textbar31 (39\textbar20)   & 38\textbar19 (6\textbar0)     & 26\textbar0 & 33\textbar0  & 209\textbar99 (65\textbar42)      \\
\textit{manban}      & 12\textbar44 (0\textbar8） & 14\textbar4 (0\textbar4)     & 35\textbar25 (32\textbar12)   & 39\textbar28 (18\textbar28)   &      &       & 100\textbar101 (50\textbar52)     \\
\textit{kuaiban}     & --          & 6\textbar0            & --              & 50\textbar29           & --   & 58\textbar0  & 114\textbar29              \\
\textit{sanyan}      & 7\textbar6         & --             & 17\textbar0            & --              & --   & 5\textbar0   & 29\textbar6                \\
\textit{kuaisanyan}  & 16\textbar16       & --             & 22\textbar13 (22\textbar11)   & --              & 10\textbar0 & --    & 48\textbar29 (22\textbar11)       \\
\textit{zhongsanyan} & --          & --             & 0\textbar4             & --              & --   & --    & 0\textbar4                 \\
\textit{mansanyan}   & --          & --             & 6\textbar6 (6\textbar6)       & --              & --   & --    & 6\textbar6 (6\textbar6)           \\
\textit{erliu}       & --          & 39\textbar12 (7\textbar12)   & --              & 22\textbar84           & --   & 8\textbar0   & 69\textbar96 (7\textbar12)        \\
\textit{liushui}     & --          & 85\textbar45 (61\textbar45)  & --              & 32\textbar34 (32\textbar23)   & --   & 17\textbar0  & 134\textbar79 (93\textbar68)      \\
\textit{daoban}      & 1\textbar0         & 2\textbar0            & 2\textbar0             & 4\textbar2             & 1\textbar0  & 5\textbar0   & 15\textbar2                \\
\textit{sanban}      & --          & 15\textbar0           & 1\textbar0             & 1\textbar9 (0\textbar1)       & 2\textbar0  & 7\textbar0   & 26\textbar9 (0\textbar1)          \\
\textit{yaoban}      & 1\textbar0         & 2\textbar1 (0\textbar1)      & 1\textbar0 (1\textbar0)       & 21\textbar12           & --   & 9\textbar0   & 34\textbar13 (1\textbar1)         \\
\textit{huilong}     & 1\textbar0         & --             & 4\textbar0             & --              & 1\textbar0  & --    & 6\textbar0                 \\
Total       & 38\textbar93 (0\textbar8) & 221\textbar84 (88\textbar84) & 142\textbar79 (100\textbar49) & 207\textbar217 (56\textbar52) & 40\textbar0 & 142\textbar0 & 792\textbar473 (293\textbar185)   \\
\bottomrule
\end{tabular}
\end{table*}

\subsection{Coverage, completeness, quality and reusability}
As stated previously, the purpose of the jingju a cappella singing audio dataset is to offer a comprehensive and complete resource for the study of automatic jingju singing evaluation as described in section \ref{sec:intro}. Based on this purpose, we evaluate four criteria for corpus creation - coverage, completeness, quality and reusability, as defined by \cite{Serra2014Creating}.

\textbf{Coverage}: The dataset includes the three main role-types - \textit{laosheng}, \textit{dan} and \textit{jing}. For \textit{laosheng} and \textit{dan} role-types, both professional and amateur singings have been recorded. 96 \textit{laosheng} and 105 \textit{dan} melodic lines in the dataset were sung both by professional and amateur singers, which allows us to analyze vocal techniques between the professional and amateur versions of the same melodic lines, and build computational similarity models by using these lines. The dataset also includes the two main \textit{shengqiang} - \textit{xipi} and \textit{erhuang}, and a few auxiliary ones, such as \textit{sipingdiao}, \textit{nanbangzi}, where the information of the auxiliary \textit{shengqiang} is not presented in table \ref{tab:lines}. 
In terms of \textit{banshi}, the whole range of metered ones is represented in the dataset - \textit{yuanban}, \textit{manban}, \textit{kuaiban}, \textit{erliu}, \textit{liushui}, \textit{sanyan} and its three variations. Besides these metered \textit{banshi}, there are a few auxiliary ones, whose occurrence is very punctual in performance. The thorough coverage of the most common \textit{shengqiang}, \textit{banshi} allows to train a singing evaluation model with good generalization ability.

\textbf{Completeness}: The dataset contains the metadata of the recordings and annotations both at the recording and the line level, organized in separate spreadsheets. For the recordings, the metadata contains the title of the work in Chinese, role-type, \textit{shengqiang}, \textit{banshi}, whether it contains \textit{jinghu} accompaniment. As for the lines, each of them is annotated with the role-type, \textit{shengqiang}, \textit{banshi}, line type, that is, opening or closing, the lyrics for the whole line and the related score in the score collection \cite{Repetto2017Score}.

\textbf{Quality}:
We conducted a small number of the recordings in room 1 and these recordings contain medium room reverberation and minor background noise. However, apart from those, the other recording sessions have been done in room 2 and those recordings are dry, clean and of good quality.

\textbf{Reusability}: The a cappella singing and \textit{jinghu} accompaniment audio recordings, their metadata are available online for research. Due to copyright issues, the commercial accompaniment audio recordings are available on request. All the audio and metadata files in this dataset are licensed under Creative Commons Attribution-NonCommercial 4.0 International. 

\subsection{Integration in the corpus}
The scores in our corpus have been gathered with the purpose of studying jingju singing regarding its musical system elements \cite{Repetto2017Score}. The a cappella audio recordings and the scores in our corpus are related through the melodic line which both of them represent. As described in \cite{Repetto2017Score}, scores and recordings are not directly related, however, the melodic contour and the lyrics are common in the majority of the pieces. Taking into account these considerations, 257 of the 705 lines (36.31\%) for \textit{laosheng} and 180 of the 512 lines (35.16\%) are related between the score collection and the a cappella singing audio dataset. The related scores are vital in the automatic singing evaluation algorithm since the score-informed method is used for the syllable segmentation step (figure \ref{fig:system_framework}).

\subsection{Potential of the dataset}

Apart from the great potential for automatic singing evaluation, the dataset allows many other sorts of musicological research. Since the dataset contains partial recordings of the \textit{jinghu} accompaniment, they are a useful resource for the analysis of the performing interaction between the singing and \textit{jinghu} lines. Some \textit{laosheng} and \textit{dan} arias were sung by both female and male singers, and they will be of benefit for analyzing differences in male and female timbre in these role-types. The scores and recordings which share lines allow a combined analysis, such as linguistic tone and melody relationship analysis. 
Finally, the audio recordings along with their annotations also will be beneficial for some basic MIR research tasks on jingju singing, such as melody extraction, structural segmentation, key detection and audio-to-lyrics alignment. 

\section{Conclusions}
In this paper, we have presented a jingju a cappella singing audio dataset for the study of automatic jingju singing evaluation. This dataset 
presents both professional and amateur singings and the most common \textit{shengqiang}, \textit{banshi} and role-types. It has been integrated into our existing CompMusic jingju corpus. Some potential usages of this dataset apart from the automatic singing evaluation have been discussed. The audio dataset, together with its annotated metadata per aria and melodic line are openly avaiable online.

In the future work, we intend to extend the dataset and increase the shared melodic lines between professional and amateur singings by recording other amateur singers. At the same time, we will exploit the potential of this dataset by annotating it in terms of melodic line and syllable boundaries, then retrain the phoneme acoustic model and syllable segmentation model presented in section \ref{sec:existing datasets}. Finally, we plan to conduct perceptual experiments to measure the similarities between professional and amateur singing syllables. These similarities along with the audio recordings will be used as the training dataset to build automatic jingju singing evaluation models.

\begin{acks}
This research was funded by the European Research Council under the European Union’s Seventh Framework Program, as part of the CompMusic project (ERC grant agreement 267583). We are thankful to WANG Xin for providing the recording equipment and to LIU Yiting for providing the recording room in NACTA.
\end{acks}

\bibliographystyle{ACM-Reference-Format}
\bibliography{sigproc} 

\end{document}